\documentclass[showpacs, secnumarabic, amssymb, nobibnotes, aps, prb, reprint]{revtex4-1}
\usepackage{graphicx}%
\usepackage{dcolumn}%
\usepackage{bm}%
\usepackage{lineno}
\usepackage{color}

\begin{document}

\title{Point defects and p-type conductivity in Zn$_{1\textrm{-}x}$Mn$_{x}$GeAs$_{2}$}

\author{L.~Kilanski}
 \altaffiliation[Also at ]{Institute of Physics, Polish Academy of Sciences, Al. Lotnikow 32/46, 02-668 Warsaw, Poland}
 \email{kilan@ifpan.edu.pl}
\author{C.~Rauch}
\author{F.~Tuomisto}
\affiliation{Department of Applied Physics, Aalto University, P.O.Box 11100, FI-00076 Aalto Espoo, Finland}

\author{A.~Podg\'{o}rni}
\author{E.~Dynowska}
\author{W.~Dobrowolski}
\affiliation{Institute of Physics, Polish Academy of Sciences, Al. Lotnikow 32/46, 02-668 Warsaw, Poland.}

\author{I.~V.~Fedorchenko}
\author{S.~F.~Marenkin}
\affiliation{Kurnakov Institute of General and Inorganic Chemistry RAS, 119991 Moscow, Russia}

\date{\today}

\begin{abstract}

Positron annihilation spectroscopy is used to study point defects in Zn$_{1\textrm{-}x}$Mn$_{x}$GeAs$_{2}$ crystals with low Mn content 0$\,$$\leq$$\,$$x$$\,$$\leq$$\,$0.042 with disordered zincblende and chalcopyrite structure. The role of negatively charged vacancies and non-open-volume defects is discussed with respect to the high $p$-type conductivity with carrier concentration 10$^{19}$$\,$$\leq$$\,$$n$$\,$$\leq$$\,$10$^{21}$$\;$cm$^{-3}$ in our samples. Neutral As vacancies, together with negatively charged Zn vacancies and non-open-volume defects with concentrations around 10$^{16}$-10$^{18}$$\;$cm$^{-3}$, increasing with the amount of Mn in the alloy, are observed. The observed concentrations of defects are not sufficient to be responsible for the strong $p$-type conductivity of our crystals. Therefore, we suggest that other types of defects, such as extended defects, have a strong influence on the conductivity of Zn$_{1\textrm{-}x}$Mn$_{x}$GeAs$_{2}$ crystals.

\end{abstract}





\maketitle


\section{Introduction}

Diluted magnetic semiconductors (DMS's) are an intensively developed group of materials designed usually on the basis of a III-V or II-VI semiconductor matrix alloyed with transition metals or rare earth elements.\cite{Kossut93a, Matsukura02a, Dobrowolski03a} Since it is difficult to obtain room temperature carrier mediated ferromagnetism in conventional DMS's like Ga$_{1\textrm{-}x}$Mn$_{x}$As (see Ref.$\;$\onlinecite{Dietl10a}) a significant attention has been turned to more complex compounds. High $p$-type conductivity and a significant solubility of Mn ions are necessary for increasing the Curie temeprature above 300$\;$K in this class of compounds. \\ \indent Mn alloyed II-IV-V$_{2}$ chalcopyrite semiconductors became a subject of considerable interest in the last few years since room temperature ferromagnetism was found in several alloys belonging to this group.\cite{Picozzi04a, Erwin04a} It was claimed that II-IV-V$_{2}$ DMS's with a high $p$-type conductivity can show itinerant or defect mediated ferromagnetism at room temperature.\cite{Mahadevan04a} The high Curie temperature together with high $p$-type conductivity with carrier concentration $n$$\,$$>$$\,$10$^{19}$$\;$cm$^{-3}$ was recently observed in one of II-IV-V$_{2}$ representatives, i.e., Zn$_{1\textrm{-}x}$Mn$_{x}$GeAs$_{2}$ crystals with $x$$\,$$>$$\,$0.07.\cite{Novotortsev04a, Kilanski09b, Kilanski10a} In our earlier papers we showed that room temperature ferromagnetism in both Zn$_{1\textrm{-}x}$Mn$_{x}$GeAs$_{2}$ and Cd$_{1\textrm{-}x}$Mn$_{x}$GeAs$_{2}$ alloys is a result of short range magnetic interactions due to the presence of MnAs nanoclusters.\cite{Kilanski10a} Moreover, our recent study reports a significant value of the Mn-ion-conducting hole exchange integral $J_{pd}$$\,$$=$$\,$(0.75$\pm$0.09)$\;$eV for Zn$_{0.997}$Mn$_{0.003}$GeAs$_{2}$ crystal.\cite{Kilanski13a} It is hence evident, that only at low dilution limit of Mn in Zn$_{1\textrm{-}x}$Mn$_{x}$GeAs$_{2}$ one can expect to observe itinerant ferromagnetism.  \\ \indent In this paper we identify different defect states in Zn$_{1\textrm{-}x}$Mn$_{x}$GeAs$_{2}$ DMS with the use of positron annihilation spectroscopy techniques. Our studies are focused on samples with low Mn content 0$\,$$\leq$$\,$$x$$\,$$\leq$$\,$0.042, where high quality Zn$_{1\textrm{-}x}$Mn$_{x}$GeAs$_{2}$ crystals without any signatures of secondary phases can be grown. We have shown earlier (see Ref.~\onlinecite{Kilanski09a}), that in the case of the samples with $x$$\,$$>$$\,$0.05 only neutral As vacancies can be identified with the use of positron annihilation spectroscopy, and these cannot be responsible for the high $p$-type conductivity typically observed in Zn$_{1\textrm{-}x}$Mn$_{x}$GeAs$_{2}$ crystals. Since positron annihilation spectroscopy is a powerful technique for probing negatively charged defects, it is highly probable that the low quality of previously studied samples was responsible for problems with detection of cation related Zn or Ge vacancies, most probably present in the crystals with high concentrations. We found that at low dilution limit a disordered zincblende structure of Zn$_{1\textrm{-}x}$Mn$_{x}$GeAs$_{2}$ alloy is preferred, except for the sample with $x$$\,$$=$$\,$0.003, where low Mn dilution seems to stabilize the chalcopyrite structure. The presence of negatively charged ionic non-open-volume defects together with negative Zn and neutral As vacancy type defects was detected for disordered zincblende and chalcopyrite crystals, respectively. The concentrations of the observed defects are in the range of 10$^{16}$-10$^{18}$$\;$cm$^{-3}$. The observed concentration of defects is not sufficient to be responsible for the strong $p$-type conductivity of our crystals with carrier concentrations 10$^{19}$$\,$$>$$\,$$n$$\,$$>$$\,$10$^{20}$$\;$cm$^{-3}$. It is therefore likely that other types of defects, not detected by positron annihilation spectroscopy, have a strong influence on the conductivity of Zn$_{1\textrm{-}x}$Mn$_{x}$GeAs$_{2}$ crystals.

\section{Basic characterization}

We investigate bulk Zn$_{1\textrm{-}x}$Mn$_{x}$GeAs$_{2}$ crystals grown by the direct fusion method from high purity ZnAs$_{2}$, Ge, and Mn powders taken in stoichiometric ratios.\cite{Novotortsev05a} The growth was performed at a temperature of about 1200$\;$K. The Mn-doped crystals were cooled from the growth temperature down to 300$\,$K with relatively high speed (about 5-10$\,$K/s) in order to improve the homogeneity of the samples and to prevent Mn clustering and diffusion out of the crystals. The as grown ingots were cut into thin slices (typically around 1$\;$mm thick) perpendicular to the growth direction with the use of a precision wire saw. Each crystal slice was chemically cleaned, etched, and mechanically polished prior further characterization.   \\ \indent The chemical composition of the samples was determined by using energy dispersive x-ray fluorescence method (EDXRF). The typical relative uncertainty of this method is not exceeding 10\% of the calculated value of $x$. The EDXRF analysis show that our Zn$_{1\textrm{-}x}$Mn$_{x}$GeAs$_{2}$ samples have Mn content $x$ in the range of 0$\;$$\leq$$\,$$x$$\;$$\leq$$\,$0.042 (see Table$\;$\ref{tab:table1}). Within the measurement accuracy all the studied crystals preserved the correct stoichiometry of Zn:Ge:As equal to 1:1:2.

\subsection{X-ray diffraction}

\noindent High resolution x-ray diffraction method (HRXRD) was used to study the structural properties of Zn$_{1\textrm{-}x}$Mn$_{x}$GeAs$_{2}$ crystals. Measurements were done with the use of multipurpose X'Pert PRO MPD, Panalytical diffractometer (Cu K$_{\alpha1}$ radiation was used with wavelength $\lambda$$\,$=$\,$1.5406$\:$$\textrm{\AA}$) configured for Bragg-Brentano diffraction geometry and equipped with a strip detector and an incident-beam Johansson monochromator. In order to increase the quality and accuracy of the diffraction patterns the data acquisition in each measurement was done over several hours. The indexing procedure of measured diffraction patterns as well as lattice parameters calculations were performed using SCANIX 2.60PC program.\cite{Paszkowicz89a} \\ \indent The analysis of the HRXRD results shows that two cubic disordered zincblende phases with $a$$\,$$=$$\,$5.6462$\pm$0.0002$\,$$\textrm{\AA}$ and $a$$\,$$=$$\,$5.9055$\pm$0.0007$\,$$\textrm{\AA}$ are the main crystallographic phases for the pure ZnGeAs$_{2}$ crystal. The addition of a small quantity of Mn ($x$$\,$$=$$\,$0.003) to the Zn$_{1\textrm{-}x}$Mn$_{x}$GeAs$_{2}$ alloy stabilizes the tetragonal chalcopyrite structure with $a$$\,$$=$$\,$5.6751$\pm$0.0002$\,$$\textrm{\AA}$ and $c$$\,$$=$$\,$11.1534$\pm$0.0005$\,$$\textrm{\AA}$. Moreover, the presence of the zincblende phase with $a$$\,$$=$$\,$5.6471$\pm$0.0004$\,$$\textrm{\AA}$ is identified in Zn$_{1\textrm{-}x}$Mn$_{x}$GeAs$_{2}$ sample with $x$$\,$$=$$\,$0.003. Further increase of the Mn content above $x$$\,$$=$$\,$0.003 results in a change of the main crystallographic phase of the alloy back to the cubic disordered zincblende structure. The lattice parameters determined for Zn$_{1\textrm{-}x}$Mn$_{x}$GeAs$_{2}$ crystals with $x$$\,$$>$$\,$0.01 are similar to the ones reported for the pure ZnGeAs$_{2}$ sample. The coexistence of cubic disordered zincblende and tetragonal chalcopyrite structures is justified by the phase diagram of ternary Zn-Ge-As system\cite{Schon94a} in which both compounds lie on the same line connecting ZnAs$_{2}$ and Ge. It must be pointed out, that diffraction patterns for both disordered zincblende and chalcopyrite (see Ref.~\onlinecite{Schon94a}) are located very close to each other and it is possible to distinguish them only with the use of a state-of-the-art diffractometer. We want to emphasize that all our crystals have almost perfect stoichiometry of ZnGeAs$_{2}$ compound, as determined with the use of the EDXRF technique. It is hence evident, that the studied alloy is ZnGeAs$_{2}$ compound, but the presence of the cubic disordered zincblende structure is a signature of a large chemical disorder of the alloy, widely observed in ternary chalcopyrite systems,\cite{Rincon92a} reflecting a mixing of the Zn and Ge atoms in the cation sublattice.

\subsection{Hall effect}

\begin{figure}[t]
 \includegraphics[width = 0.42\textwidth, bb = -10 20 600 550]
 {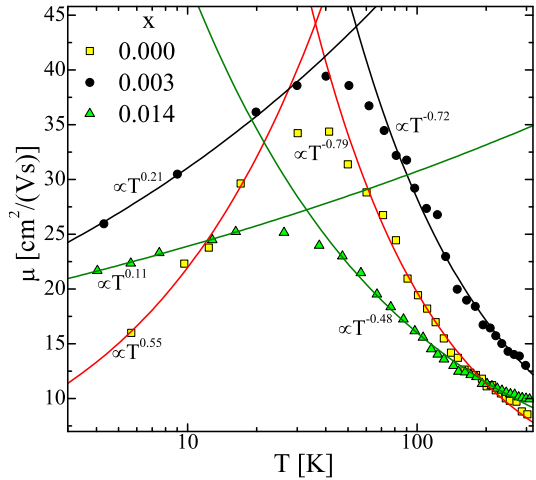}%
 \caption{\label{FignvsT} Temperature dependence of the Hall carrier mobility $\mu$ measured (points) for the Zn$_{1\textrm{-}x}$Mn$_{x}$GeAs$_{2}$ samples with different chemical content $x$. The lines represent the power functions fitted to the experimental data.}
 \end{figure}

\noindent In order to obtain information about fundamental electrical properties of the Zn$_{1\textrm{-}x}$Mn$_{x}$GeAs$_{2}$ alloy, temperature dependent magnetotransport measurements were performed. The standard six contact dc method was used for electrical characterization of the samples. The detailed magnetotransport studies of the Zn$_{1\textrm{-}x}$Mn$_{x}$GeAs$_{2}$ samples studied in this work are presented in Ref.$\;$\onlinecite{Kilanski13a}. For the purposes of this work we revisit the low-field magnetotransport results e.g. the resistivity $\rho_{xx}$ and the Hall effect measurements carried out in the temperature range from 4.3 up to 320$\;$K. The Hall effect measurements were performed at stabilized magnetic field $B$$\,$$=$$\,$$\pm$1.5$\;$T.  \\ \indent Initially, the temperature dependence of the resistivity parallel to the current direction, $\rho_{xx}$, in the absence of external magnetic field is studied. The obtained results show, that in the case of all our samples a metallic $\rho_{xx}$($T$) dependence is observed, a behavior characteristic of degenerate semiconductors. It indicates that the carrier transport is not due to thermal activation of band carriers. Resistivity values obtained at $T$$\,$$=$$\,$300$\;$K for samples with different chemical compositions are summarized in Table$\;$\ref{tab:table1}.
\begin{table}[t]
\caption{\label{tab:table1}%
Results of the basic Zn$_{1\textrm{-}x}$Mn$_{x}$GeAs$_{2}$ sample characterization including chemical content $x$ and electrical properties: resistivity $\rho_{xx}$, carrier concentration $p$ and carrier mobility $\mu$ obtained at $T$$\,$$=$$\,$300$\;$K.}
\begin{ruledtabular}
\begin{tabular}{cccc}

$x$$\pm$$\Delta x$ & $\rho_{xx}$$\pm$$\Delta$$\rho_{xx}$ &   $p$$\pm$$\Delta$$p$    & $\mu$$\pm$$\Delta$$\mu$  \\
                   &  [10$^{-2}$$\;$$\Omega$$\cdot$cm]   & [10$^{19}$$\;$cm$^{-3}$] &  [cm$^{2}$/(V$\cdot$s)]  \\ \hline

0               &  8.8$\pm$0.1 &  8.3$\pm$0.2 &  8.6$\pm$0.3  \\
0.003$\pm$0.001 & 24.9$\pm$0.1 &  1.9$\pm$0.1 & 13.0$\pm$0.4  \\
0.014$\pm$0.001 &  5.4$\pm$0.1 & 11.6$\pm$0.3 & 10.0$\pm$0.4  \\
0.027$\pm$0.002 & 14.3$\pm$0.1 &  7.6$\pm$0.2 &  5.7$\pm$0.2  \\
0.042$\pm$0.004 &  5.9$\pm$0.1 & 10.8$\pm$0.3 &  9.9$\pm$0.3  \\

\end{tabular}
\end{ruledtabular}
\end{table}
The results indicate only a small difference between the resistance values for the investigated samples. There seems to be no evident trend of the $\rho_{xx}$ values with the chemical composition of the Zn$_{1\textrm{-}x}$Mn$_{x}$GeAs$_{2}$ alloy. \\ \indent The measurements of the Hall effect as a function of temperature allow us to determine the temperature dependence of the Hall carrier concentration, $p$, in all the our samples. The results show that all the Zn$_{1\textrm{-}x}$Mn$_{x}$GeAs$_{2}$ crystals have $p$-type conductivity with relatively high carrier concentrations in the range of 10$^{19}$$\,$$\leq$$\,$$p$$\,$$\leq$$\,$10$^{20}$$\;$cm$^{-3}$ and relatively low carrier mobilities 5$\,$$<$$\,$$\mu$$\,$$<$$\,$13$\;$cm$^{2}$/(V$\cdot$s).\cite{Kilanski13a} It is generally considered that the high concentration of conducting holes in ZnGeAs$_{2}$ is due to the existence of a large number of negatively charged Zn or Ge vacancy type defects.\cite{Mercey86a} A difference of the Hall carrier concentration is observed in two Zn$_{1\textrm{-}x}$Mn$_{x}$GeAs$_{2}$ samples with similar chemical content ($x$$\,$$<$$\,$0.004) possessing chalcopyrite and disordered zincblende structure, respectively. This difference is very likely due to the fact that in a disordered zincblende sample the chemical disorder on the cation sites is much larger than in chalcopyrite structure, which results in a higher concentration of electrically active defects leading to an increase in the concentration of free conducting holes. All our zincblende crystals with different Mn content show similar carrier concentration $p$ and mobility $\mu$ suggesting that Mn alloying does not significantly affect the number of defects in the material. \\ \indent The temperature dependence of the Hall carrier mobility (Fig.$\;$\ref{FignvsT}) is an increasing function of the temperature at $T$$\,$$<$$\,$50$\;$K, while at $T$$\,$$>$$\,$50$\;$K the trend is opposite. The positive slope of $\mu$($T$) dependence at $T$$\,$$<$$\,$50$\;$K is a signature that the ionic-scattering mechanism is involved in the carrier transport at low temperatures. On the other hand, the negative slope of $p$($T$) dependence at $T$$\,$$>$$\,$50$\;$K is a signature of a phonon scattering n our material. The influence of different scattering mechanisms on the temperature dependence of the carrier mobility in a degenerate semiconductor can be expressed with the use of Matthiessen's Rule $\mu^{-1} = \mu^{-1}_{ph} + \mu^{-1}_{d_{i}}$, where $\mu^{-1}_{ph}$ is the lattice scattering due to phonons and $\mu^{-1}_{d_{i}}$ is the scattering due to $i$-th defect type present in the material. The scattering due to ionized impurities can be exactly expressed within the Brooks-Herring theory. There are two major reasons why it is not possible to apply Brooks-Herring formulation: (i) the relevant material parameters for ZnGeAs$_{2}$ are not known and (ii) the carrier mobility of the samples is well below 50$\;$cm$^{2}$/(V$\cdot$s) even at low temperatures which means that the carrier transport can not be strictly described by Drude theory.
The low temperature region ($T$$\,$$<$$\,$50$\;$K) of the $\mu$($T$) dependence for a $p$-type semiconductor is expected to be proportional to $\propto$$\,$$T^{-2.2}$ for GaAs while for our samples the exponent is much lower. It indicates that more than a single type of scattering centers are involved in the reduction of the $\mu$($T$) dependence at low temperatures. At $T$$\,$$>$$\,$50$\;$K the $\mu$($T$) dependence can be fitted to the power law with exponents in the range -0.5$\ldots$-0.8 for most of the samples. The mobility due to acoustic phonon scattering is expected to be proportional to $T^{-3/2}$, while the mobility due to optical phonon scattering only is expected to be proportional to $T^{-1/2}$. The values obtained for our samples point into the conclusion that the optical phonon scattering is the major scattering process in ZnGeAs$_{2}$ crystals.

\section{Defect identification by means of positron annihilation spectroscopy}

\subsection{Experimental details}

\noindent Positron annihilation experiments consisted of positron lifetime and Doppler broadening measurements performed on the as-grown Zn$_{1\textrm{-}x}$Mn$_{x}$GeAs$_{2}$ crystals. The temperature of the sample was controlled during the experiment with the use of a closed-cycle helium cryostat and a resistive heating system in the range of 15$\,$$\leq$$\,$$T$$\,$$\leq$$\,$520$\;$K. \\ \indent The positron lifetime was measured using a standard fast-fast coincidence spectrometer with a time resolution of about 250$\;$ps.\cite{Saarinen98a} The 20$\;$$\mu$Ci positron source ($^{22}$Na deposited on a 1.5-$\mu$m thick Al foil) was placed between two identical sample pieces during the measurement. Typically about 3$\times$10$^{6}$ annihilation events were collected for each measurement point. The lifetime spectrum expressed as a sum of exponential decays $n(t)$$\,$$=$$\,$$\sum_{i}$$I_{i}$$\exp(-t/\tau_{i})$ can be decomposed into a few components convoluted with the Gaussian resolution function of the spectrometer, after background and source component subtraction. The positron in state $i$ annihilates with a lifetime $\tau_{i}$ and intensity $I_{i}$. The state in question can be the delocalized state in the lattice or the localized state at a vacancy type defect. The average positron lifetime $\tau_{ave}$$\,$$=$$\,$$\sum_{i}$$\tau_{i}$$I_{i}$ is insensitive to decomposition procedure, and even small changes around 1$\,$ps can be reliably measured. The increase of $\tau_{ave}$ above the lifetime of a perfect crystal, $\tau_{B}$, is a signature, that vacancy type defects at which positrons are effectively trapped, are present in a material with a concentration higher than 10$^{15}$$\;$cm$^{-3}$. In the case of one dominant vacancy type defect with a specific lifetime, $\tau_{V}$, the decomposition of experimental spectra into two lifetime components, $\tau_{1}$ and $\tau_{2}$, is straightforward to interpret since $\tau_{2}$$\,$$=$$\,$$\tau_{V}$.   \\ \indent The Doppler broadening measurements were carried out simultaneously with positron lifetime measurements with the use of a high-purity Ge detector with an energy resolution of 1.3$\;$keV at 511$\;$keV. The Doppler spectra are  analyzed with the conventional $S$ and $W$ parameters, defined as fractions of counts in the low momentum range ($p_{z}$$\,$$<$$\,$0.4$\;$a.u.) and high momentum range (1.6$\;$a.u.$\,$$<$$\,$$p_{z}$$\,$$<$$\,$4$\;$a.u.), respectively. Typically, the electron-positron momentum distribution narrows ($S$ parameter increases) when positrons annihilate as trapped at vacancies. The $W$ parameter is more sensitive than the $S$ parameter to the chemical identities of the atoms surrounding the positron annihilation site, as core electrons have wider momentum distributions compared to less localized valence electrons.

\subsection{Results and analysis}

\begin{figure}[t]
 \includegraphics[width = 0.42\textwidth, bb = 0 0 369 750]{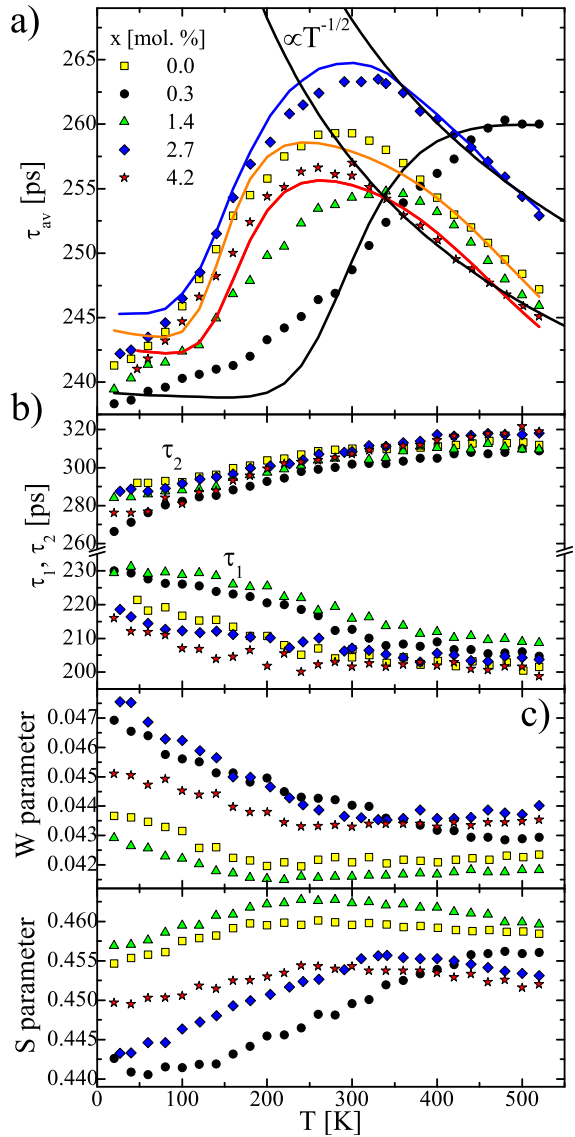}%
 \caption{\label{FigTauAvvsT} The temperature dependencies of the positron lifetime and positron Doppler broadening data including: (a) the average positron lifetime measured (points) and fitted to the positron trapping model (lines) (b) positron lifetime spectra decomposition results including  the lifetimes $\tau_{1}$ and $\tau_{2}$ and (c) the Doppler broadening $S$ and $W$ parameters obtained for the Zn$_{1\textrm{-}x}$Mn$_{x}$GeAs$_{2}$ samples containing different amount of Mn (see legend).}
\end{figure}

Temperature dependent positron lifetime measurements were performed in order to identify defect types and their charge state in Zn$_{1\textrm{-}x}$Mn$_{x}$GeAs$_{2}$ crystals (see Fig.$\;$\ref{FigTauAvvsT}).
The obtained $\tau_{ave}$ is higher in the entire studied temperature range than the bulk lifetime $\tau_{B}$$\,$$=$$\,$225$\pm$5$\;$ps estimated earlier for the perfect ZnGeAs$_{2}$ lattice.\cite{Kilanski09a} This indicates that positrons are trapped at vacancy defects in these samples. \\ \indent However, the $\tau_{ave}$($T$) dependencies for disordered-zincblende and chalcopyrite structured Zn$_{1\textrm{-}x}$Mn$_{x}$GeAs$_{2}$ samples have different shapes. A decrease of $\tau_{ave}$ for $T$$\,$$<$$\,$300$\;$K down to $\tau_{ave}$(20$\,$K)$\,$$\approx$240$\pm$5$\,$ps is due to positron trapping in negatively charged ion-type defects, present in disordered-zincblende samples. The presence of non-open-volume ionic traps is further justified by the fact that $\tau_{ave}$ decreases at low temperatures to the value close to $\tau_{B}$$\,$$=$$\,$225$\pm$5$\;$ps. The shape of the $\tau_{ave}$($T$) curves at 50$\,$$<$$\,$$T$$\,$$<$$\,$270$\;$K is a typical signature of positron detrapping from these ionic traps. The decrease of the average positron lifetime with increasing temperature at $T$$\,$$>$$\,$350$\;$K is a clear indication of the observed vacancy defects being in the negative charge state. These temperature dependencies originate from the $T^{-1/2}$ dependence of the trapping coefficient for negatively charged defects, and thermal escape from shallow (less than 150$\;$meV positron binding energy) levels produced by negative ion-type defects.\cite{Saarinen98a} \\ \indent In the case of the chalcopyrite Zn$_{1\textrm{-}x}$Mn$_{x}$GeAs$_{2}$ sample the $\tau_{ave}(T)$ dependence shows a positive slope for $T$$\,$$<$$\,$450$\;$K and a saturation at higher temperatures. There are two possible explanations for the difference between the temperature dependence for this sample compared to the rest. In principle, the binding energy to the negative ion type defects could be clearly higher in this sample, resulting in the efficient detrapping only at higher temperatures, as observed in GaN (Ref.$\;$\onlinecite{Tuomisto07a}). However, in this case the data at low temperatures (at $T$$\,$$<$$\,$200$\;$K) should saturate to a constant value. Hence the second explanation is more likely, i.e., that the vacancy defects observed in this sample are in the neutral charge state, allowing for the negative ions to affect the lifetime data at much higher temperatures (since the trapping to neutral vacancies is temperature-independent). \\ \indent A general increase of the average positron lifetime (compared at $T$$\,$$=$$\,$300$\:$K) with increasing Mn content is observed. It suggests that the concentration of vacancy type defects is an increasing function of Mn content in the samples. In order to make more detailed analysis of the defect identities and concentrations, the different components of the spectra need to be resolved. \\ \indent The temperature dependent positron lifetime data can be described within simple positron trapping model.\cite{Saarinen98a} In this model, the trapping coefficient $\mu_{V}$$\,$$=$$\kappa_{V}$/$c_{V}$ to electrically neutral vacancies does not depend on temperature. On the other hand, in the case of negatively charged vacancies the trapping coefficient varies as $\mu_{V}$$\,$$\propto$$\,$$T^{-1/2}$. Positrons can be also trapped at hydrogen-like Rydberg states surrounding negatively charged ion-type defects with trapping rate varying also as $\mu_{V}$$\,$$\propto$$\,$$T^{-1/2}$ (for details see Ref.~\onlinecite{Puska90a}). The trapping of positrons to a shallow traps is especially important at low temperatures. The thermal escape of positrons from hydrogen-like Rydberg states can be described using Eq.$\;$\ref{Eq01}
\begin{equation}\label{Eq01}
    \delta_{st}=\mu_{R}\bigg{(}\frac{m^{*}_{+}k_{B}T}{2 \pi \hbar^{2}}\bigg{)}^{\frac{3}{2}}\times\exp(-E_{b,st}/k_{B}T),
\end{equation}
where $\mu_{R}$ is the positron trapping coefficient to the lowest hydrogen-like Rydberg state, $E_{b,st}$ is the positron binding energy of the lowest Rydberg state, and $m^{*}_{+}$$\,$$\simeq$$\,$$m_{0}$ is the effective mass of the positron. In principle, positrons may also escape from the Rydberg states around negatively charged vacancies, but we assume that this transition is fast enough so that this effect can be neglected. The above assumption is supported by results obtained in the case of GaN (Ref.~\onlinecite{Saarinen98a}) and other compound semiconductors. An effective trapping rate of the shallow traps can be expressed with the use of the following equation
\begin{equation}\label{Eq02}
    \kappa_{st}^{eff}= \frac{\kappa_{st}}{1+\delta_{st}/\lambda_{st}},
\end{equation}
where $\lambda_{st}$$\,$$\simeq$$\,$$\lambda_{B}$ is the annihilation rate of positrons trapped at the Rydberg state, and $\kappa_{st}$$\,$$=$$\,$ $\mu_{R}$$V_{st}$ is directly related to the concentration of negatively charged ionic positron traps $c_{st}$. \\ \indent The positron lifetime spectra can be usually decomposed into a few lifetime components related to positron annihilation in different states. The average lifetime is expressed as
\begin{equation}\label{Eq03}
    \tau_{av}=\eta_{B}\tau_{B}+\sum_{j}\eta_{D_{j}}\tau_{D_{j}},
\end{equation}
where $\eta$ and $\tau$ are the annihilation fraction and positron lifetime in the free state of the lattice B and $j$-th defect state D. The annihilation fractions are related to the trapping rates through the following equations
\begin{equation}\label{Eq04}
    \eta_{B}=\frac{\lambda_{B}}{\lambda_{B}+\sum_{j}\kappa_{D_{j}}^{eff}},  \qquad \eta_{D_{j}}=\frac{\kappa_{D_{j}}^{eff}}{\lambda_{B}+\sum_{j'}\kappa_{D_{j'}}^{eff}}
\end{equation}
\\ \indent The positron lifetime spectra are decomposed into two lifetime components, $\tau_{1}$ and $\tau_{2}$. Fitting of more than two components results in a very large statistical error, indicating that only two components can be resolved. The temperature dependencies of $\tau_{1}$ and $\tau_{2}$ and the second lifetime intensity, $I_{2}$, are presented in Fig.$\;$\ref{FigTauAvvsT}b. At the highest measurement temperatures the estimated bulk lifetime $\tau_{B}$$\,$$=$$\,$($\tau_{1}$$\cdot$$\tau_{2}$)/($\tau_{1}$+$\tau_{2}$-$\tau_{ave}$)$\,$$=$$\,$235$\pm$10$\;$ps. This is close to the earlier estimated bulk lifetime of 225$\;$ps. The increase of $\tau_{1}$ (towards $\tau_{B}$) with decreasing temperature is a result of lifetime mixing caused by the trapping of positrons by negative ions that produce the same lifetime as the defect-free lattice. If negative ions were not present, the shorter component $\tau_{1}$ would decrease with decreasing temperature as the trapping to the negatively charged vacancy defects increases ($\tau_{1}$$\,$$=$$\,$$(\tau_{B}^{-1} + \kappa)^{-1}$ in the single-defect model). \\ \indent The measurements of the Doppler broadening of the electron-positron annihilation line were performed simultaneously with the positron lifetime measurements as shown in Fig.$\;$\ref{FigTauAvvsT}. The temperature dependence of the Doppler broadening $S$ and $W$ parameters shows a qualitatively similar behavior as that of the average positron lifetime, but as opposed to the lifetime behavior, the changes with temperature are clearly smaller than the variations from sample to sample. Hence the effects of the negatively charged vacancies and negative ion-type defects are clearly observable, but the $S$ and W parameters suggest that the exact defect identities may be different in differently Mn-doped samples.

\section{Discussion}

\subsection{Positron lifetimes}

\noindent The results of the present studies extend our previous investigations of point defects in Zn$_{1\textrm{-}x}$Mn$_{x}$GeAs$_{2}$ crystals with rather high Mn content (0.053$\,$$\leq$$\,$$x$$\,$$<$$\,$0.182) (see Ref.$\;$\onlinecite{Kilanski09a}) into low Mn dilution limit 0$\,$$\leq$$\,$$x$$\,$$\leq$$\,$0.042. There are several differences between the current samples and the previously studied crystals, mostly in their structural properties.\cite{Kilanski10a} In contrast to the previously studied samples, the aggregation of magnetic ions into MnAs clusters is not observed, hence the positron annihilation measurements are not affected by the positron trapping into metallic precipitates. In addition it was possible to produce crystals of two different crystal structures, i.e., disordered zincblende and chalcopyrite structure, which allows us to analyze the effect of the crystal structure on the defect generation. \\ \indent
Equation (\ref{Eq03}) can be fitted to the temperature dependent positron lifetime spectra (Fig.$\;$\ref{FigTauAvvsT}a) using the trapping rates and binding energies to the Rydberg states as the fitting parameters. As we can see there we obtained rather fair agreement between the experimental results and the fitted theoretical lines. It is a clear signature, that the defect identification was not complete in the studied material. The best fits to the experimental data were obtained with the positron binding energy values close to $E_{b,st}$$\,$$\simeq$$\,$90$\pm$15$\;$meV and $E_{b,st}$$\,$$\simeq$$\,$170$\pm$10$\;$meV for disordered zincblende and chalcopyrite samples, respectively. The concentrations of defects can be calculated assuming the trapping coefficient to the negatively charged vacancies to be around $\mu_{Vk}$$\,$$=$$\,$3$\times$10$^{15}$$\;$s$^{-1}$ at 300$\;$K for cation defect,\cite{Puska90a} the concentration of negative vacancies in disordered zincblende crystals was equal to [$V_{D}$]$\,$$=$$\,$($\kappa_{D}$/$\mu_{D}$)$\cdot$$N_{at}$$\,$$\simeq$$\,$2.8$\div$4.4$\times$10$^{16}$$\;$cm$^{-3}$, where  $N_{at}$$\,$$=$$\,$4.2$\times$10$^{22}$$\;$cm$^{-3}$ is the atomic density of ZnGeAs$_{2}$ compound. \\ \indent The concentration of negative vacancies observed herein is an increasing function of Mn content $x$. On the other hand, in the case of neutral anion vacancy defects observed in chalcopyrite crystal the trapping coefficient is lower than in the case of  disordered zincblende samples and equal to $\mu_{Va}$$\,$$=$$\,$1$\times$10$^{15}$$\;$s$^{-1}$.\cite{Puska90a} The concentration of neutral vacancy type defects estimated in the case of chalcopyrite sample was equal to [$V_{D}$]$\,$$=$$\,$1.7$\times$10$^{17}$$\;$cm$^{-3}$. Additionally, the concentrations of ionic non-open-volume defects is estimated from the trapping coefficient $\kappa_{st}$ to be close to
[$V_{st}$]$\,$$=$$\,$2.5$\times$10$^{17}$$\;$cm$^{-3}$ in the disordered zincblende crystals and [$V_{st}$]$\,$$=$$\,$7.5$\times$10$^{17}$$\;$cm$^{-3}$ for the chalcopyrite sample. \\ \indent The concentration of both ionic non-open-volume and vacancy type defects, obtained on the basis of positron data in the case of all the studied crystals are insufficient to be able to be the source of high concentrations of conducting holes $n$$\,$$\simeq$$\,$10$^{19}$$\div$10$^{20}$$\;$cm$^{-3}$. However, in the case of the chalcopyrite crystal, the presence of rather high concentration of negative defects is detected, contrary to the results presented in Ref.$\;$\onlinecite{Kilanski09a}. This is probably due to much better structural quality of the present samples, with respect to older samples. \\ \indent The higher lifetime component $\tau_{2}$ is an increasing function of temperature (see Fig.$\;$\ref{FigTauAvvsT}b) in all the studied samples up to temperatures around 400$\;$K, where it saturates with values around $\tau_{2}$$\,$$\simeq$$\,$315$\pm$5$\;$ps. This could mean that positrons are trapped into two different types of vacancy defects in all investigated crystals. In the case of GaAs, the Ga sublattice vacancy has a lifetime of about 260$\;$ps, and the As vacancy about 295$\;$ps in the neutral and about 260$\;$ps in the negative charge state.\cite{Saarinen98a} Since the bulk lifetime observed in the case of Zn$_{1\textrm{-}x}$Mn$_{x}$GeAs$_{2}$ crystals is similar to that of GaAs, and the higher lifetime component $\tau_{2}$ is similar to that of neutral As vacancy in GaAs, it would be probable that in the present case we do again observed the presence of As vacancies, similarly as in our previous work (Ref.$\;$\onlinecite{Kilanski09a}). The mixing of two different types of defects into $\tau_{2}$ is consistent with the fits of the temperature dependent average positron lifetime reproducing the experimental data with moderate accuracy. \\ \indent The ternary ZnGeAs$_{2}$ compound, an isoelectronic analogue of binary GaAs, is normally obtained as $p$-type material with carrier concentration $\sim$10$^{18}$..10$^{19}$$\;$cm$^{-3}$ at room temperature.\cite{Brudnyi1983a} The control their electrical properties is difficult. It is believed that the high $p$-type conductivity in this material is stimulated by the presence of $V_{Zn}$ and $Ge_{As}$ defects.\cite{Brudnyi1983a} It is therefore tempting to assign the experimentally observed defects to these.

\subsection{Doppler broadening data}

\begin{figure}
 \includegraphics[width = 0.42\textwidth, bb = 0 16 600 540]{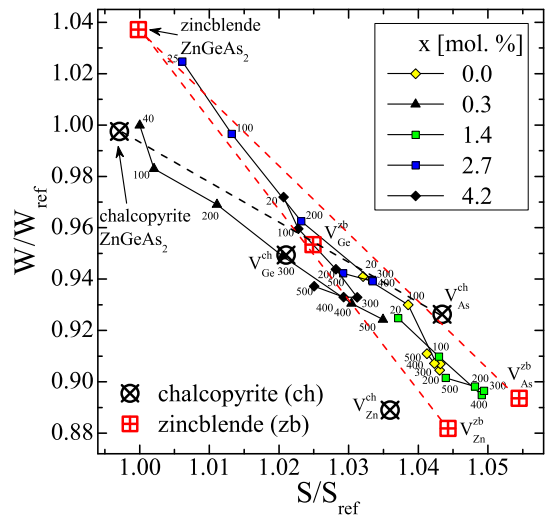}%
 \caption{\label{FigSWPlot} Characteristic $W$ vs $S$ parameters (normalized to the lowest experimental values $S_{ref}$$\,$$=$$\,$0.440 and $W_{ref}$$\,$$=$$\,$0.047) obtained at different temperatures (see labels) for selected Zn$_{1\textrm{-}x}$Mn$_{x}$GeAs$_{2}$ samples with different chemical compositions (see legend) together with the theoretical values of $W$ vs $S$ calculated for the free state of the lattice and for different defects: zinc $V_{Zn}$, germanium $V_{Ge}$, and arsenic $V_{As}$ monovacancies for the  chalcopyrite and zincblende structure.}
 \end{figure}

A typical form of presenting results of the positron Doppler broadening spectroscopy, enabling a more detailed interpretation of the data, is plotting them on the $W$($S$) plane. In order make the $W$($S$) plot more clearly legible only some selected points obtained at temperatures separated by 100$\;$K are selected and plotted for all the Zn$_{1\textrm{-}x}$Mn$_{x}$GeAs$_{2}$ samples shown in Fig.$\;$\ref{FigSWPlot}.
The experimentally observed positron Doppler broadening parameters, namely $S$ and $W$, are modeled theoretically using ab-initio methods.\cite{Makkonen06a} The zero positron density limit of the two component density functional theory (TCDFT) (Ref.~\onlinecite{Boronski86a}) is used to calculate structural and electronic properties of the chalcopyrite and zincblende Zn$_{1\textrm{-}x}$Mn$_{x}$GeAs$_{2}$ crystals. A local density approximation (LDA) together with a projector augmented-wave method (PAW) implemented in Viena Ab-initio Simulation Package (VASP)\cite{Kresse96a} are used during a self-consistent calculations of the valence electron densities in the studied structures. \\ \indent The calculations of electronic properties of both chalcopyrite and zincblende ZnGeAs$_{2}$ crystals are done using 64-atom supercell structures. The chemical disorder present in disordered zincblende samples is neglected in the present calculations. The estimated cutoff energy for the studied system is equal to 320~eV. The calculations are done with 3s and 4p electrons of Zn, Ge, and As treated as valence electrons. The lattice and ionic relaxation is considered before electronic structure calculations with a convergence criterium for forces of maximum 0.01$\;$eV/$\textrm{\AA}$ by including positron induced forces on ions. During the electronic structure calculations the Brillouin zone is sampled with 3$^{3}$ Monkhorst-Pack $k$-point mesh. \\ \indent The positron states are calculated using the so-called conventional scheme, and the LDA and state-dependent scheme\cite{Alatalo96a} are used for the calculation of positron annihilation rates and the description of many-body effects in the calculation of the momentum distributions of annihilating electron-positron pairs. The final momentum distributions are convoluted with a gaussian function of 0.53 a.u. FWHM to simulate the experimental resolution in common coincidence Doppler measurements. All momentum distributions are calculated along the [001] crystal axis. The values of the $S$ and $W$ parameters calculated for the free state in the chalcopyrite and zincblende lattices as well as for Zn ($V_{Zn}^{ch}$, $V_{Zn}^{zb}$), Ge ($V_{Ge}^{ch}$, $V_{Ge}^{zb}$), and As ($V_{As}^{ch}$, $V_{As}^{zb}$) monovacancies are gathered in Fig.$\;$\ref{FigSWPlot} together with the experimental data.  The calculated parameters for the defects and disordered zincblende bulk are normalized to the calculated values of chalcopyrite bulk. \\ \indent Defects with no open volume, such as negative ionic impurities, produce the same annihilation parameters as the bulk lattice and thus are not observed in the slopes of the ($S$,$W$) plot. Points on the $S$($W$) plane obtained for the lowest temperatures  for the 0.3\% and 2.7\% samples, located in the left-upper corner of Fig.$\;$\ref{FigSWPlot} are close to the values characteristic for the annihilation of positrons in the volume of the crystal, namely ($S_{B}$;$W_{B}$) for chalcopyrite and disordered zincblende crystals, respectively. \\ \indent The temperature dependent $W$($S$) curve obtained for chalcopyrite Zn$_{0.997}$Mn$_{0.003}$GeAs$_{2}$ sample (marked with triangles in Fig.$\;$\ref{FigSWPlot}) forms a nearly straight line lying close to the line connecting the characteristic positron Doppler parameters for the free state of the lattice and two defect types: (i) negative Zn ($V_{Zn}^{ch}$) and neutral As vacancies $V_{As}^{ch}$), respectively. The presence of two different types of open-volume defects is also visible in $\tau_{2}$($T$) dependence indicating that at high temperatures ($T$$\,$$>$$\,$450$\;$K) the positron trapping into the neutral As vacancies is dominant while at low temperatures ($T$$\,$$<$$\,$50$\;$K) the trapping of positrons to the negatively charged ionic defects is accompanied by positron trapping into negative Zn vacancies. For the zincblende Zn$_{1\textrm{-}x}$Mn$_{x}$GeAs$_{2}$ samples the observed $W$($S$) dependencies point into similar intepretation as for the case of chalcopyrite crystal. The calculated values of the $W$($S$) points for different types of point defects connected to bulk parameters $W_{B}$($S_{B}$) form lines with similar slopes. The experimental points lie on the $W$($S$) plane away from the line connecting ZnGeAs$_{2}$ lattice and $V_{Ge}^{zb}$. It indicates that the positron trapping is negligible in this type of negatively charged defect and it is likely that the concentration of germanium vacancies in our samples is below 10$^{15}$$\;$cm$^{-3}$. The $W$($S$) experimental points for disordered zincblende Zn$_{1\textrm{-}x}$Mn$_{x}$GeAs$_{2}$ samples lie within a triangle connecting the points characteristic of ZnGeAs$_{2}$ lattice, Zn and As vacancies. It is a signature that the positron trapping at $T$$\,$$>$$\,$270$\;$K occur at both Zn and As defect types.

\section{Summary}

We applied positron annihilation spectroscopy to study point defects in as-grown bulk Zn$_{1\textrm{-}x}$Mn$_{x}$GeAs$_{2}$ crystals. The present investigations are focused over low Mn-dilution limit of ZnGeAs$_{2}$ with chemical composition of the crystals varying in the range of 0$\,$$\leq$$\,$$x$$\,$$\leq$$\,$0.042. The samples crystallized in cubic disordered zincblende or tetragonal chalcopyrite structures without any signatures of Mn-clustering. Hole conductivity with high carrier concentration 10$^{19}$$\,$$\leq$$\,$$n$$\,$$\leq$$\,$10$^{20}$$\;$cm$^{-3}$ and low mobility $\mu$$\,$$<$$\,$35$\;$cm$^{2}$/(V$\cdot$s) was observed. An increase in the carrier concentration accompanied with a decrease in mobility when increasing the Mn amount in the samples indicated poor Mn allocation in the alloy causing an increase of charged defect concentration with $x$. \\ \indent The temperature dependent positron lifetime and Doppler broadening spectroscopies showed the presence of more than one dominant defect in Zn$_{1\textrm{-}x}$Mn$_{x}$GeAs$_{2}$ crystals. At low temperatures, $T$$\,$$<$$\,$150$\;$K, positrons were trapped into two types of defects: (i) negatively charged non-open-volume ionic type defects, arising from structural disorder, having concentrations of an order of 2.5$\times$10$^{17}$$\,$$\leq$$\,$[$V_{st}$]$\,$$\leq$$\,$7.5$\times$10$^{17}$$\,$cm$^{-3}$, and (ii) negatively charged Zn vacancies, with concentration increasing as a function of Mn content in the alloy. At higher temperatures the neutral As vacancies were observed in all the samples with concentrations of about 10$^{17}$$\;$cm$^{-3}$. Despite the detection of a complex defect structure present in the samples, the estimated concentrations of point defects are still unable to fully explain the high $p$-type conductivity of the Zn$_{1\textrm{-}x}$Mn$_{x}$GeAs$_{2}$ crystals. It is therefore very likely that extended defects or other kinds of extended structures that are not observed with positron annihilation spectroscopy are present in the material and increase the concentration of conducting holes in this material.

\section{Acknowledgments}

\noindent  Scientific work was financed from funds for science in 2009-2013, under the project no. IP2010017770 granted by Ministry of Science and Higher Education of Poland and project no. N$\,$N202$\,$166840 granted by National Center for Science of Poland, and by the Academy of Finland. We acknowledge the Aalto University Science-IT project for computing resources. This work has been supported by the RFBR Project No. 13-03-00125 and Russian Federation Project No. MK-1454.2014.3. We thank Dr. I.~Makkonen for helpful discussions.


\begin{thebibliography}{natbib}

\bibitem{Kossut93a} J.~Kossut and W.~Dobrowolski, Handbook of Magnetic Materials (North-Holland, Amsterdam, 1993), Vol. 7, pp. 231–305.

\bibitem{Matsukura02a} F.~Matsukura, H.~Ohno, and T.~Dietl, Handbook of Magnetic Materials (Elsevier, Amsterdam, 2002), Vol. 14, Chaps. III–V, pp. 1–87.

\bibitem{Dobrowolski03a} W.~Dobrowolski, J.~Kossut, and T.~Story, Handbook of Magnetic Materials (Elsevier, New York, 2003), Vol. 15, Chaps. II–VI and IV–VI, pp. 289–377.

\bibitem{Dietl10a} T.~Dietl, \emph{Nature Materials} \textbf{9}, 965 (2010).

\bibitem{Picozzi04a} S.~Picozzi, \emph{Nature Materials} \textbf{3}, 349 (2004).

\bibitem{Erwin04a} S.~C.~Erwin and I.~\v{Z}uti\'{c}, \emph{Nature Materials} \textbf{3}, 410 (2004).

\bibitem{Mahadevan04a} P.~Mahadevan and A.~Zunger, \emph{Phys. Rev. Lett.} \textbf{88}, 047205 (2002).

\bibitem{Novotortsev04a} V.~M.~Novotortsev, S.~F.~Marenkin, S.~A.~Varnavskii, L.~I.~Koroleva, T.~A.~Kupriyanova, R.~Szymczak, L.~Kilanski, and B.~Krzymanska, \emph{Russian Journal of Inorganic Chemistry} \textbf{53}, 22 (2008).

\bibitem{Kilanski09b} L.~Kilanski, M.~G\'{o}rska, V.~Domukhovski, W.~Dobrowolski, J.~R.~Anderson, C.~R.~Rotundu, S.~A.~Varniavskii, and S.~F.~Marenkin, \emph{Acta Phys. Pol. A} \textbf{114}, 1151–1157 (2008).

\bibitem{Kilanski10a} L.~Kilanski, M.~G\'{o}rska, W.~Dobrowolski, E.~Dynowska, M.~W\'{o}jcik, B.~J.~Kowalski, J.~R.~Anderson, C.~R.~Rotundu, D.~K.~Maude, S.~A.~Varnavskiy, I.~V.~Fedorchenko, and S.~F.~Marenkin, \emph{J. Appl. Phys} \textbf{108}, 073925 (2010).

\bibitem{Kilanski13a} L.~Kilanski, K.~Sza\l{}owski, R.~Szymczak, M.~G\'{o}rska, E.~Dynowska, P.~Aleshkevych, A.~Podg\'{o}rni, A.~Avdonin, W.~Dobrowolski, I.V.~Fedorchenko, and S.F.~Marenkin, \emph{J. Appl. Phys.} \textbf{114}, 093908 (2013).

\bibitem{Kilanski11a} L.~Kilanski, W.~Dobrowolski, E.~Dynowska, M.~W\'{o}jcik, B.~J.~Kowalski, N.~Nedelko, A.~\'{S}lawska-Waniewska, D.~K.~Maude, S.~A.~Varnavskiy, I.~V.~Fedorchenko, S.~F.~Marenkin, \emph{Solid State Commun.} \textbf{151}, 870 (2011).

\bibitem{Kilanski09a} L.~Kilanski, A.~Zubiaga, F.~Tuomisto, W.~Dobrowolski, V.~Domukhovski, S.~A.~Varnavskiy, and S.~F.~Marenkin, \emph{J. Appl. Phys.} \textbf{106}, 013524 (2009).

\bibitem{Novotortsev05a} V.~M.~Novotortsev, V.~T.~Kalinnikov, L.~I.~Koroleva, R.~V.~Demin, S.~F.~Marenkin, T.~G.~Aminov, G.~G.~Shabunina, S.~V.~Boichuk, and V.~A.~Ivanov, \emph{Russ. J. Inorg. Chem.} \textbf{50}, 492 (2005).

\bibitem{Paszkowicz89a} W.~J.~Paszkowicz, \emph{J. Appl. Crystallogr.} \textbf{22}, 186 (1989).

\bibitem{Schon94a} S.~Sch\"{o}n, M.~L.~Fearheiley, K.~Diesner, and S.~Fiechter, \emph{J. Cryst. Growth} \textbf{135}, 601 (1994).

\bibitem{Rincon92a} C.~Rinc\'{o}n, \emph{Phys. Rev. B} \textbf{45}, 12716 (1992).

\bibitem{Mercey86a} B.~Mercey, D.~Chippaux, J.~Vizot, and A.~Deschanvres, \emph{J. Phys. Chem. Solids} \textbf{47}, 37 (1986).

\bibitem{Saarinen98a} F.~Tuomisto and I.~Makkonen, \emph{Rev. Mod. Phys.} \textbf{85}, 1583 (2013).

\bibitem{Tuomisto07a} F.~Tuomisto, V.~Ranki, D.~C.~Look, and G.~C.~Farlow, \emph{Phys. Rev B} \textbf{76}, 165207 (2007).

\bibitem{Tuomisto05a} F.~Tuomisto, K.~Saarinen, D.~C.~Look, and G.~C.~Farlow, \emph{Phys. Rev. B} \textbf{72}, 085206 (2005).

\bibitem{Puska90a} M.~J.~Puska, C.~Corbel, and R.~M.~Nieminen, \emph{Phys. Rev. B} \textbf{41}, 9980 (1990).

\bibitem{Brudnyi1983a} V.~N.~Brudnyi, A.~I.~Potapov, and Yu.~V.~Rud, \emph{Phys. Stat. Sol. (a)} \textbf{75}, K73 (1983).

\bibitem{Makkonen06a} I.~Makkonen, M.~Hakala, and M.~J.~Puska, \emph{Phys. Rev. B} \textbf{73}, 035103 (2006).

\bibitem{Boronski86a} E.~Boro\'{n}ski and R.~M.~Nieminen, \emph{Phys. Rev. B} \textbf{34}, 3820 (1986).

\bibitem{Kresse96a} G.~Kresse and J.~Furthm\"{u}ller, \emph{Phys. Rev. B} \textbf{54}, 11169 (1996).

\bibitem{Alatalo96a} M.~Alatalo, B.~Barbiellini, M.~Hakala, H.~Kauppinen, T.~Korhonen, M.~J.~Puska, K.~Saarinen, P.~Hautoj\"{a}rvi, and R.~M.~Nieminen, \emph{Phys. Rev. B} \textbf{54}, 2397 (1996).




\end{thebibliography}
\end{document}